# Observation of Blood Flow in Major Neck Vessels Modulated by Physiological Maneuvers

Gennadi Saiko[1*], Timothy Burton[2], Faraz Sadrzadeh-Afsharazar[2], Shota Yamashita[3], Kenshin Shimono[3], Yasuyuki Kakihana[3], and Alexandre Douplik[1,4]

1   Department of Physics, Toronto Metropolitan University, Toronto, Canada
2   Department of Biomedical Engineering, Toronto Metropolitan University, Toronto, Canada
3   Department of Emergency and Intensive Care Medicine, Kagoshima University, Kagoshima, Japan
4   iBEST, Keenan Research Centre of the Li Ka Shing (LKS) Knowledge Institute, St.Michael Hospital, Toronto, Canada
*   Correspondence: gsaiko@torontomu.ca;

Abstract: Large neck vessels (carotid artery and internal jugular vein, IJV) offer a unique opportunity to monitor hemodynamics non-invasively by optical means. The primary shortcoming of past work has been the focus on healthy volunteers in normal physiological conditions and well-controlled environments. To drive the technology closer to the bedside, testing is required under more realistic conditions, including in pathologies and real-world environments (e.g., similar to ICU or emergency care settings). The primary goal of the current work was to extend the range of physiological maneuvers for blood flow modulation by introducing new maneuvers and observing PPG response to them. The data from the necks of two healthy volunteers in a supine position were collected by clinical PPG and in-house built PPG sensors, accompanied by ECG signal collection. Seven maneuvers (abdominojugular test, breath holding, Valsalva, proximal occlusion of right IJV, distal occlusion of right IJV, proximal occlusion of left IJV, distal occlusion of left IJV) were performed in sequence with 1 min allocated for each maneuver. The 1 min was split into three segments: baseline (15 s), experiment (15 s), and recovery (30 s). Thus, the overall duration of the experiment was 7 min. AC amplitude from clinical PPG, DC amplitudes from in-house built PPG, and ECG signal were compared during all seven physiological maneuvers. Newly proposed maneuvers (Valsalva and IJV occlusions) demonstrated modulation of blood flow, which was more significant than previously reported maneuvers (abdominojugular test and breath holding). The proposed physiological maneuvers demonstrate high potential as instruments for modulating blood flow in major neck vessels.

1. Introduction
Large neck vessels (for example, carotid artery, CA, internal jugular vein, IJV) provide a unique opportunity to investigate the cardiovascular system. In particular, they are a) relatively close to the heart, b) quite large, c) quite shallow (1-2cm in depth), and d) easily accessible. Moreover, they supply and drain the brain and can be used to study cerebral circulation. As a result, the carotid artery and internal jugular vein are commonly used in multiple medical procedures and research.

In addition, this unique combination provides an excellent opportunity to monitor hemodynamics non-invasively. As a result, the carotid artery is a frequent target for multiple ultrasound investigations. Similarly, the internal jugular vein is used in several clinical tests, including the evaluation of heart function.

With advances in optics and instrumentation, the large neck vessels became accessible by optical means, which triggered the recent influx of research in this area. One area of such research is an investigation of the waveforms caused by blood propagation in the carotid artery and internal jugular vein. These waveforms are deemed to be an indicator of the blood pressure in the corresponding vessel.

In particular, several noncontact imaging methods have been proposed to capture and investigate pressure waveforms, including PPG imaging by a color camera[1], nonuniform light by a color camera[2], subpixel image registration[3], and Specular Reflection Vascular Imaging (SRVI)[4].

In addition to remote PPG modalities, the contact PPG (cPPG) was tried successfully to extract jugular vein pressure (JVP) waveforms from the anterior jugular vein[5].

The primary shortcoming of previous works is their focus on healthy volunteers in normal physiological conditions and well-controlled environments. To drive the technology closer to the bedside, it needs to be tested in more realistic conditions, which include pathological conditions and an environment that approaches real-world (e.g., similar toICU or emergency care settings).

Investigation of pathologies in real clinical settings presents a significant challenge, as the availability of patients with certain pathologies can be a challenge in many cases. In addition, the scale of such pathology can be unknown, and just pure investigation for scientific reasons can be a significant and unethical burden to a patient. Thus, modulating the physiological response in healthy volunteers to mimic these pathologies with physiological maneuvers is beneficial. This approach offers several advantages: controlled magnitude, required scale, and reproducibility.

In the previous article, we proposed using abdominojugular testing and breath-holding for the modulation of venous pressure, and we observed changes in blood flow in major neck blood vessels caused by this modulation[6]. However, the primary shortcoming of these maneuvers is related to the magnitude of blood flow changes they cause. They appeared to be subtle. Adding other maneuvers, which cause stronger body response, would be advantageous.

To observe blood flow changes, we explored contact and noncontact (remote) photoplethysmography (PPG) [6]. While noncontact PPG demonstrated results that align with contact PPG ones, the practicality of noncontact PPG in emergency care settings is quite questionable. In particular, the methodology is very sensitive to motion artifacts and ambient light, which are difficult to control, particularly in field settings. On the other hand, contact PPG may overcome these challenges. In particular, if the contact sensor is small and flexible enough, it may move with the skin, thus mitigating motion artifacts. Similarly, the sensor's case may shield the photodetector from the ambient light, thus reducing its impact. Therefore, we focus on the contact PPG modality in the current work.

Also, in the previous round of experiments, we observed several problems related to a clinical PPG sensor. While it provides valuable information as a pulse oximeter, this information is not sufficient for comprehensive analysis. In practice, the clinical PPG sensor provides a blood oxygenation reading and a single waveform. However, this waveform is a filtered signal processed using a highpass filter. As such, only information about cardiac cycle changes is retained, while the important information about the slowly changing component (DC component) is removed. However, this DC component may bring valuable information about the total amount of blood in vessels. The second disadvantage of the clinical PPG is a fixed (and quite small) sampling rate (250 samples per second). In some scenarios (for example, measuring pulse transit time), such a small resolution (4 ms) can be insufficient.

As such, we decided to develop an in-house built PPG sensor, where we have more control over collected data.

Thus, the aim of this preliminary work is two-fold. The primary goal was to broaden the spectrum of physiological maneuvers by exploring additional physiological manipulations and capturing the body's response using PPG. The secondary goal was to try an in-house built sensor and compare it with the clinical PPG.

2. Materials and Methods

The new physiological maneuvers were introduced and tried on healthy volunteers in addition to the previously reported maneuvers [6]. The data were collected by clinical PPG and in-house built PPG.

2.1. In-house Sensor

The in-house built PPG system is depicted in Fig 1. It consists of 2 LEDs (660 and 850nm) and a photodiode powered by Arduino microcontroller. The source/detection separation can be adjusted. It was set to 2cm in this round of experiments. The collected data are streamed to PC, where they are logged for further data processing. The sensor allows data collection for two channels at 400 samples per second.

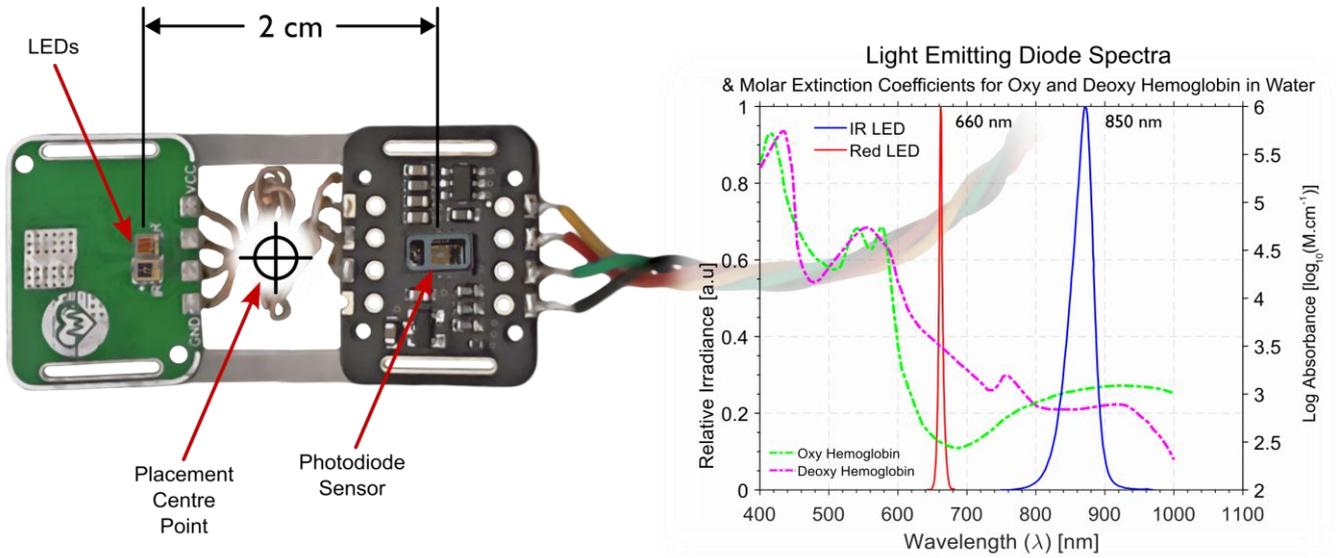

Figure 1. In-house built PPG sensor (left pane) and its spectral characteristics (right pane). For convenience, emission bands are superimposed on absorption spectra of oxy- and deoxyhemoglobins.

2.2. Data Collection

The data were collected from two healthy volunteers. To collect data, each volunteer was placed in a supine position. The clinical PPG sensor was placed on the right side of the neck over IJV. The in-house built sensors were placed on the left side of the neck over IJV. The location of IJV was verified using vascular ultrasound.

Both clinical PPG and in-house built sensors were placed across IJV (source-detector separation is perpendicular to the IJV direction).

In addition, the one-lead ECG was collected. ECG and the clinical PPG sensor, placed on the right side of the neck, were plugged into a clinical monitor. As a result, the time series from ECG and clinical PPG were synchronized. After experiments, ECG and PPG time series were extracted for further data processing.

The in-house-built sensor collected data and streamed it to a personal computer, which logged them for further data processing.

2.3. Maneuvers

In the previous article, we used two physiological maneuvers to modulate venous pressure: abdominojugular testing and breath holding [6]. However, the effect of these maneuvers appeared quite subtle. Thus, we decided to extend the range of maneuvers by adding the Valsalva maneuver and Internal Jugular Vein occlusions.

2.3.1. Valsalva maneuver

The Valsalva maneuver is forced expiration against a closed glottis. Performing the Valsalva maneuver causes an increase in intrathoracic pressure, reducing preload to the heart[7]. As a result, baroreflex and other compensatory reflex mechanisms initiate a series of cardiovascular changes. According to[7], the Valsalva maneuver can be divided into four phases, two on the onset and two on the release of strain:

"Phase I, which corresponds to the onset of strain, is associated with a transient rise in blood pressure because of the emptying of some blood from the large veins and pulmonary circulation.

Phase II follows this when positive intrathoracic pressure leads to a reduced venous return to the heart. Because of reduced venous return and thus reduced preload, stroke volume falls; this leads to a fall in blood pressure activating the baroreceptors in the carotid sinus and aortic arch. The vagal withdrawal followed by increased sympathetic discharge ensues, leading to marked tachycardia, increased cardiac output, and vasoconstriction, which leads to the recovery of blood pressure to normal values in healthy individuals.

Phase III is the transient phase involving the release of strain which leads to a sudden dip in blood pressure. The release of positive pressure leads to expansion of the pulmonary vascular bed and reduces the left ventricular cross-sectional area resulting in a transient fall in blood pressure.

Phase IV is the overshoot of the blood pressure above the baseline, which is because of the resumption of normal venous return to the heart stimulated by the sympathetic nervous system during Phase II. The overshoot of blood pressure leads to stimulation of baroreflex, leading to bradycardia and the return of blood pressure to the baseline."

Valsalva maneuver is used for the diagnostic of numerous conditions, including autonomic function assessment[8], heart failure assessment[9], diagnostics of murmurs[10], and termination of arrhythmias.

2.3.2. Internal jugular vein occlusion

Internal jugular vein occlusion aimed to explicitly modulate blood flow through the internal jugular vein. As the PPG and in-house sensors were placed on different sides of the neck, the occlusion was applied sequentially at two points (below and above the sensor) on each side. Thus, four occlusions were performed in total. The occlusion was performed by applying pressure by a finger over the internal jugular vein. The location of IJV was verified using vascular ultrasound and marked by a marker.

The reasoning for such maneuvers is as follows. If the sensor is obstructed from below (proximal to the heart), it is isolated from the vena cava and heart. However, it remains sensitive to blood flow from the brain. Similarly, if the sensor is obstructed from above (distal with respect to the heart), then it is isolated from blood flow from the brain. However, it remains sensitive to blood volume in the vena cava and heart. Moreover, tentatively, such a configuration is supposed to be sensitive to any blood flow redistribution between veins if such redistribution occurs. In particular, both arterial and venous systems have some redundancies. For example, the brain's arteries are arranged into the circle of Willis, which is believed to create redundancy for collateral circulation in the cerebral circulation. Similarly, one could expect redundancy in cerebral venous circulation as well. For example, it is known that the redundancy of the collateral venous system (except for the Rolandic vein[11]) prevents circulatory disturbances if major collecting veins are occluded. Anastomotic channels are the primary source of this redundancy. In particular, many anastomotic channels exist between the superficial venous systems, veins at the base of the brain, and transcerebral veins[12]. As a result, the primary cerebral drainage mechanism depends on numerous factors, including posture. In particular, the IJV is thought to be the principal outflow pathway for intracranial blood in the supine position. In the seated and erect positions, the IJV collapses, shifting blood flow to the vertebral veins with contribution from the spinal epidural veins, condylar veins, as well as occipital and mastoid emissary veins that connect the posterior cranial fossa dural venous sinuses to the vertebral venous system[13], [14].

Nevertheless, despite this complexity, it is reasonable to expect that the occlusion of one branch of the primary draining pathway (which is IJV in the supine position) may trigger noticeable changes in the second branch.

2.4. Protocol

All seven maneuvers (abdominojugular test (AJT), breath holding (BH), Valsalva, proximal occlusion of right IJV, distal occlusion of right IJV, proximal occlusion of left IJV, distal occlusion of left IJV) were performed in sequence with 1 min allocated for each maneuver. The 1 min was split into three segments: baseline (15 s), experiment (15 s), and recovery (30 s). Thus, the overall duration of the experiment was 7 min.

2.5. Data Processing

The data from the ECG, clinical PPG and in-house sensor was first temporally segmented into ranges based on the timing of the maneuvers. Once separated, three individual analyses were performed to assess the impact of the maneuvers: clinical PPG waveform morphology, clinical PPG AC amplitude, and novel sensor DC amplitudes.

As distal occlusion of left IJV (LS Upper IJV) signal was corrupted in one occurrence, the whole LS Upper IJV data were removed from analysis, which left six maneuvers (abdominojugular test (AJT), breath holding (BH), Valsalva, proximal occlusion of right IJV, distal occlusion of right IJV, proximal occlusion of left IJV).

Clinical PPG waveform morphology was computed using landmarks from the simultaneously-acquired ECG, which possesses the characteristic PQRST waveform, originating from atrial depolarization (P), ventricular depolarization (QRS) and ventricular repolarization (T). The fiducial point that can be found most reliably is the R-peak, which is maximal ventricular depolarization. The Pan-Tompkin algorithm[15] is a robust tool for R-peak detection, and likely the most widely used, and thus was adopted in the present workflow. The R-peaks identified by Pan-Tompkin were used to define cardiac cycles using conservative static intervals: 300ms backwards in time from the R-peak to encompass the P-wave, and 700ms forwards in time to encompass the T-wave. 300ms is more than sufficient to capture the average PR interval (from the onset of the P wave to the R peak) of 150ms[16], in addition to the average QR duration with of 50ms [17](half of the typical 100ms, given the symmetry of the QRS). Similarly, in the forward direction, the average QT interval is approximately 350-400ms[18], including the already accounted for QR duration of 50ms, resulting in a range of 300-350 ms, which is well encompassed within the 700ms allowance. Thus, a significant portion of the isoelectric period is also captured.

Once the cardiac cycle was identified, the timepoints were applied to the clinical PPG. Repeated over all the cardiac cycles, the result was a matrix of clinical PPG cycles. A dimensionality reduction and clustering methodology was then applied to select the most representative cardiac cycle. Specifically, the time dimensionality of the matrix was reduced using principal component analysis such that each cardiac cycle was represented in two dimensions. The algorithm DBSCAN (i.e., Density-Based Spatial Clustering of Applications with Noise[19]) grouped the cycles into clusters, allowing for the segregation of any outliers or otherwise unusual cycles. The largest cluster by number of cycles was selected as representative, from which, the specific cycle with minimum distance to the other cluster members was chosen for visualization.

With respect to the other analyses, AC amplitude was computed using a dynamic envelope approach [4]: the median R-R interval was calculated, as using to define an upper envelope (moving k-point maximum) and a lower envelope (moving k-point minimum). The median difference between the envelopes was calculated as the amplitude. The AC amplitude during the maneuver was compared using a ratio with the mean amplitudes during baseline and recovery. Finally, the DC change in the novel sensor was assessed by calculating the mean DC across the last second of the maneuver as compared to the first second, expressed as a ratio.

3. Results

To analyze data we performed several comparisons.

3.1. AC Component

To analyze AC component, we used "envelope" approach. The amplitude across all maneuvers extracted from clinical PPG data is depicted in Figure 2.

a)

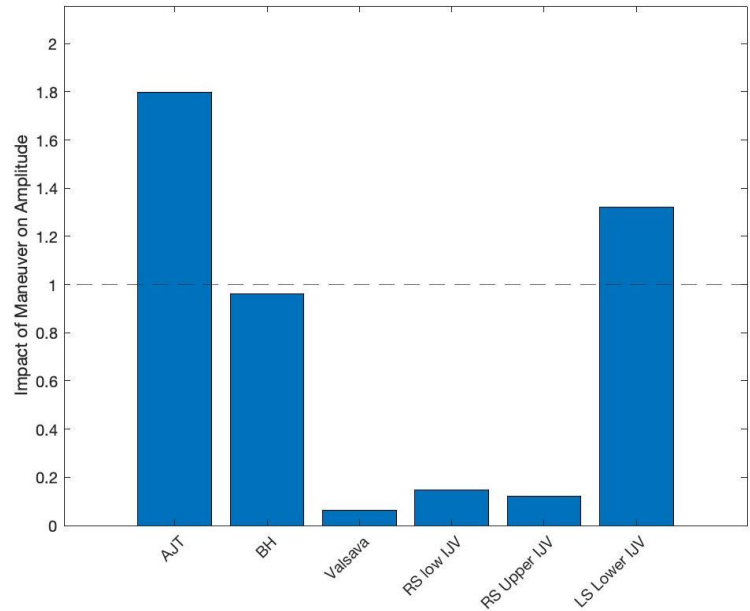

b)

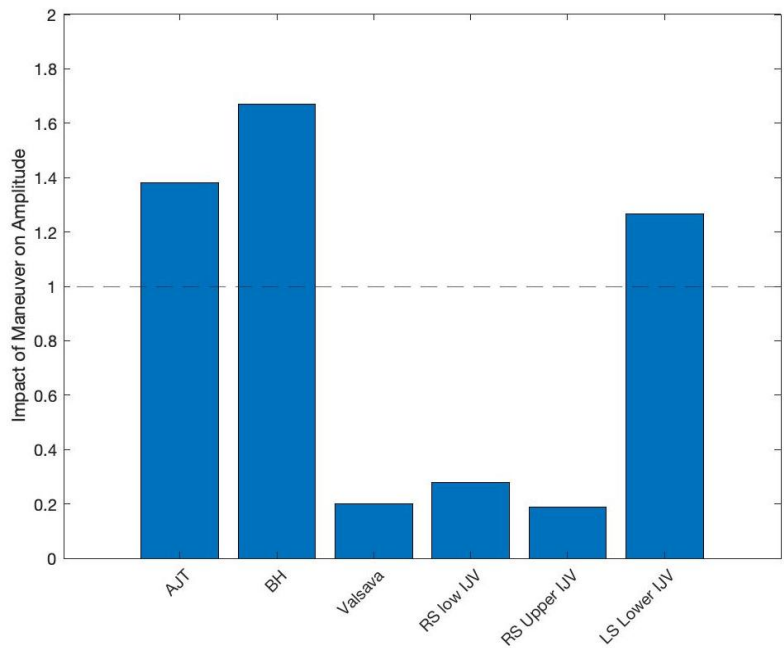

Figure 2. The effect of the maneuvers on pulsations extracted from clinical PPG data for volunteer 1 (a) and volunteer 2 (b), represented as the ratio of the amplitude during the maneuvers over the average during baseline and recovery. The maneuvers are ordered temporally, from abdominojugular test (AJT), breath holding (BH), Valsava, proximal

occlusion of right internal jugular vein (IJV), distal occlusion of right IJV and proximal occlusion of left IJV. Due to a data capture error, distal occlusion of left IJV is not available.

3.2. DC Component

As the clinical PPG sensor includes filtering, the DC component cannot be analyzed. As such, only DC components extracted from in-house built sensor were analyzed. They are displayed in Figure 3.

a)
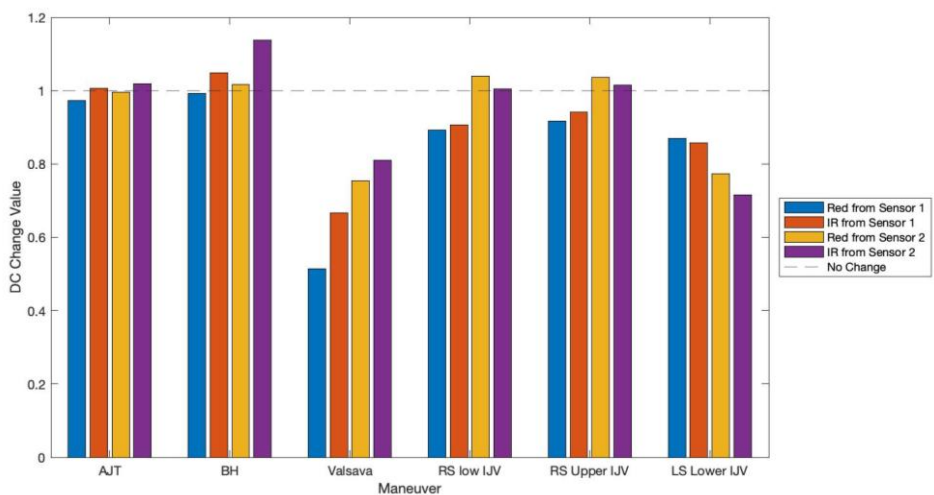

b)
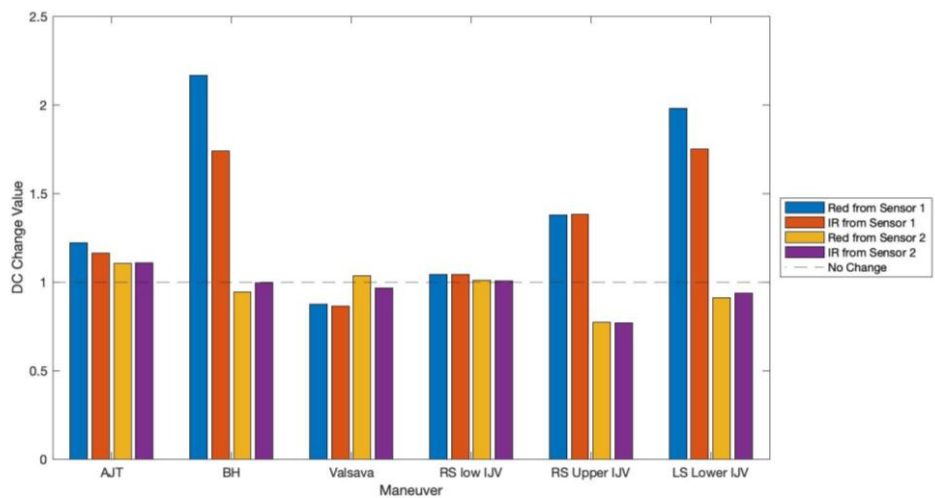

Figure 3. The DC components of PPG signal extracted from the in-house built PPG sensor for volunteer 1 ( a) and volunteer 2 (b), represented as the change in amplitude from the first second of the maneuvers to the last second. The maneuvers are ordered temporally, from abdominojugular test (AJT), breath holding (BH), Valsava, proximal occlusion of right internal jugular vein (IJV), distal occlusion of right IJV and proximal occlusion of left IJV. Due to a data capture error, distal occlusion of left IJV is not available.

3.3. Representative Waveforms

As we aimed to analyze the precise shape of the signal, synchronization with ECG is required. As such, only clinical PPG data was analyzed. The representative PPG waveforms selected by DBSCAN algorithm extracted from clinical PPG data is depicted in Fig 3 for volunteer 1, and Fig 4 for volunteer 4.

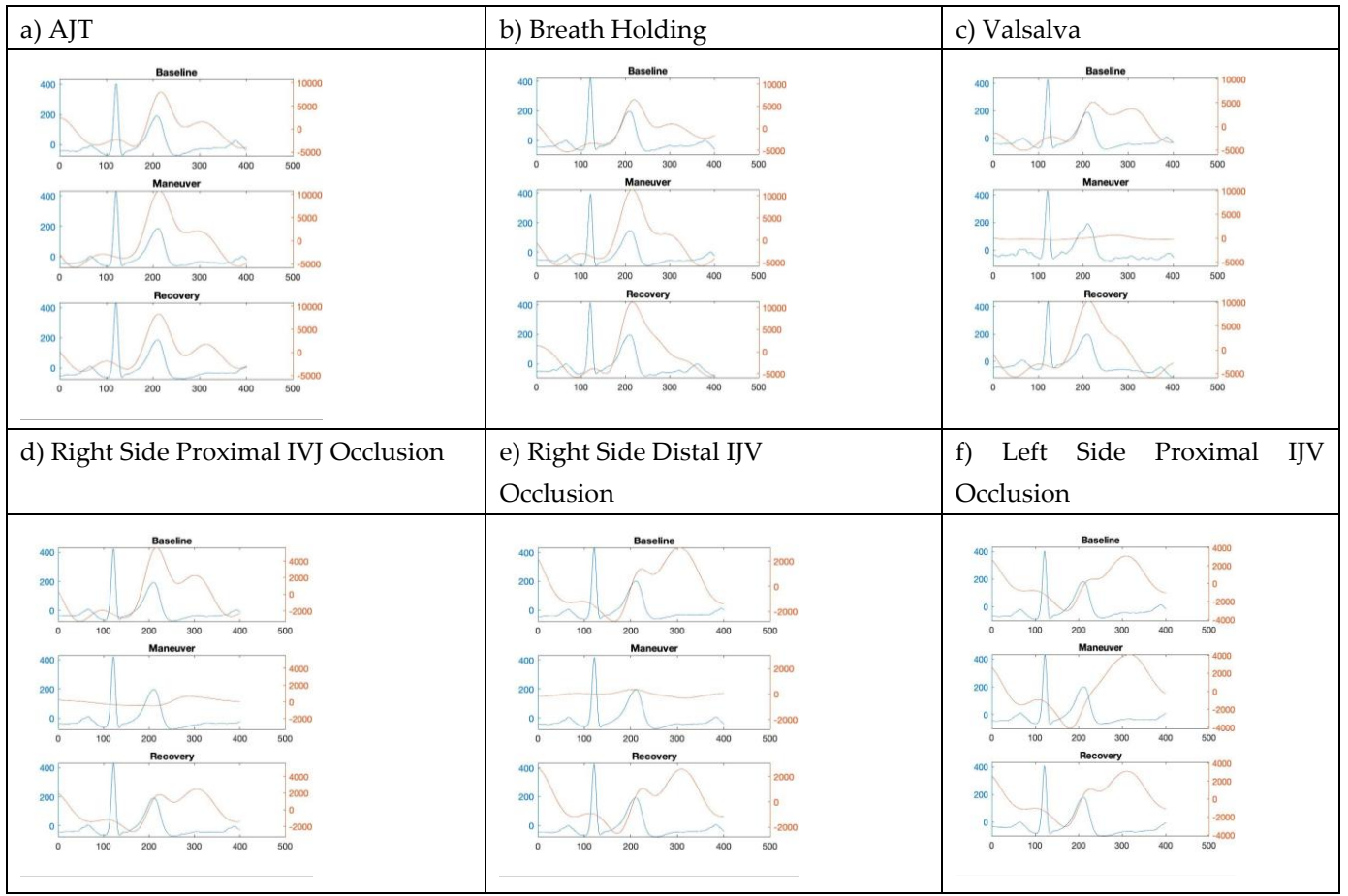

Figure 4. The representative PPG waveforms extracted from clinical PPG data for volunteer 1. Each plot consists of ECG data and PPG data at baseline and during maneuver. The clinical PPG is shown in red, and the ECG is shown in blue.

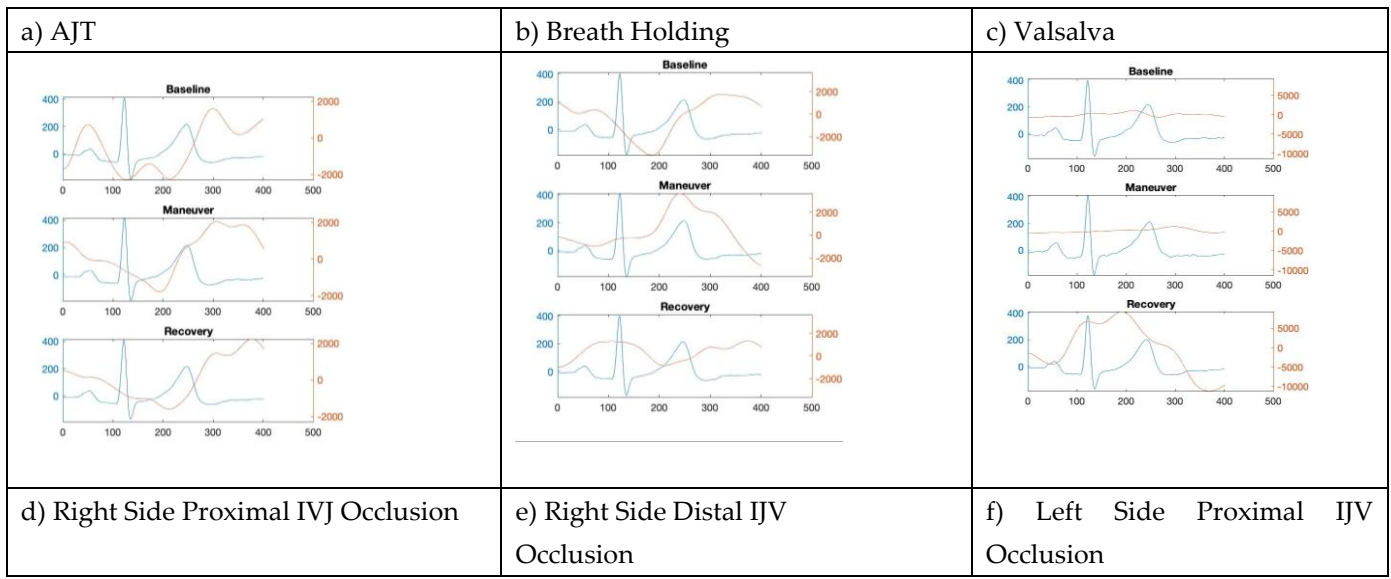

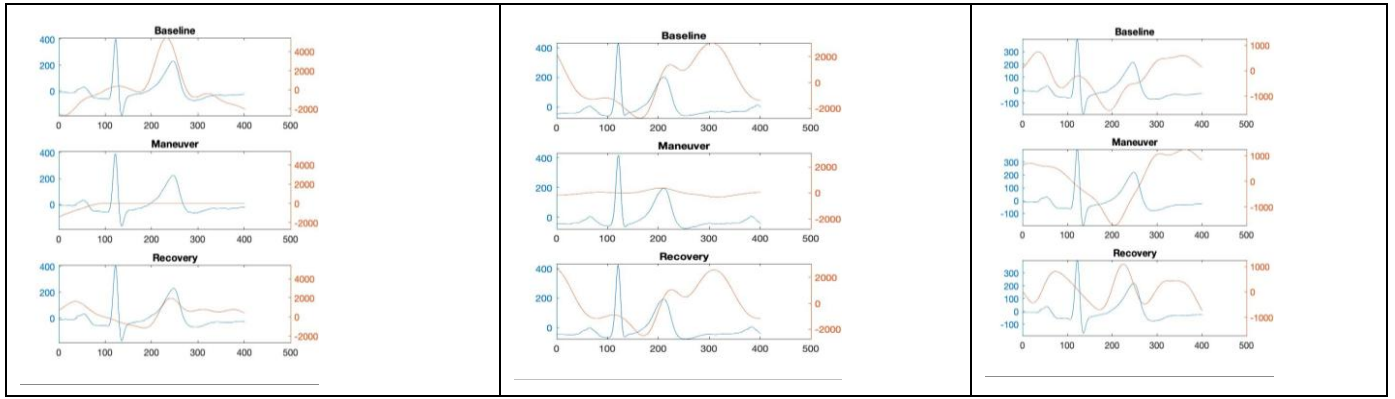
Figure 5. The representative PPG waveforms extracted from clinical PPG data for volunteer 2. Each plot consists of ECG data and PPG data at baseline and during maneuver. The clinical PPG is shown in red, and the ECG is shown in blue.

4. Discussion

The proposed maneuvers caused blood flow modulation in large neck vessels, which were detected by optical sensors. Both clinical PPG and in-house-built PPG sensors were able to identify changes.

Changes were noticeable in the PPG signal's AC and DC components (see Fig 2 & 3).

The captured waveforms have complex shapes (See Fig 4& 5). Even though the PPG sensors were placed across IJV, the origin of the obtained waveforms is unclear. While they have a complex shape, the dominant peak (see Fig 4) was typically close to the T-wave of the ECG signal, which may indicate its carotid artery origin. However, it can be a mixture of arterial and venous waveforms. Thus, in such collection geometry, the sensor potentially collects signals from both the carotid artery and the internal jugular vein. As such, further investigation is required.

Out of six physiological maneuvers, Valsalva demonstrated the most profound and consistent effect on the collected signal. Firstly, it is characterized by a significant decrease in the amplitude of pulsations. Both sensors detected this phenomenon. Secondly, it displayed the drop in the DC values measured by the in-house built sensor. The reason for this drop in the DC component value is unclear. However, most likely, it is caused by the significant distension of blood vessels. It can also explain the decrease in AC components, as their relative changes become smaller in magnitude with the distension of blood vessels. However, further investigation of this phenomenon (e.g., using vascular ultrasound) is required.

The occlusions of the internal jugular vein caused the decrease in AC amplitude on the same side as the occlusion (see Fig 2). However, on the DC component, the internal jugular vein occlusions caused less consistent results (see Fig 3). It can partially be explained that there are several mechanisms of cerebral drainage, and the interplay between them is not well understood. In general, cerebral venous circulation comprisessupratentorial venous drainage by the superficial and deep cerebral veins, including the subependymal and medullary veins, and infratentorial drainage via the bridging veins thatdrainthe posterior fossa. Both pathways eventually lead into the dural venous sinuses, which subsequently drain into the IJV in the supine position and primarily into thevertebral and paraspinous veinsin the upright position[20]. In addition, some venous drainage occurs through the extracranial veins via emissary veins, which, according to some research, can account for 30-40% of overall cerebral drainage[21].With such plurality of drainage pathways, it is not surprising that the effect of IJV occlusion on blood flow is not straightforward. Further investigation of this phenomenon is warranted.

The primary shortcoming of the current work is the small sample size (just two participants), which does not allow extrapolating any findings. In future work, we aim to validate our findings on a larger set of volunteers. Also, we plan to compare our findings with those of vascular ultrasound, which is the de facto gold standard in the field.

Another shortcoming was the lack of data synchronization across all modalities. While the ECG and clinical PPG data were synchronized, the in-house built PPG sensor was not synchronized with the ECG, which complicated their comparison. We plan to address data synchronization in the next round of experiments.

5. Conclusions

Spatially resolved contact PPG was able to capture changes in neck vessels caused by physiological maneuvers. Newly proposed maneuvers (Valsalva and IJV occlusions) demonstrated modulation of blood flow, which was more significant than previously reported maneuvers (abdominojugular test and breath holding). The proposed physiological maneuvers demonstrate high potential as instruments for modulating blood flow in major neck vessels. Further research on a larger set of subjects and comparison with existing clinical modalities is warranted.

Author Contributions: Conceptualization, G.S and Y.K.; methodology, G.S. and Y.K.; software, T.B.; hardware, F.S.A.; formal analysis, T.B. and G.S.; resources, Y.K.; data curation, T.B.; writing—original draft preparation, G.S.; writing—review and editing, T.B. and Y.K.; visualization, T.B.; supervision, A.D.; project administration, A.D.; funding acquisition, A.D. All authors have read and agreed to the published version of the manuscript.

Funding: This research was funded by NSERC I2I Grant (G.S and A.D.) and personal NSERC Discovery Grants (G.S. and A.D.)

Institutional Review Board Statement: Ethical review and approval were waived for this study due to exception for case studies with two or less participants

Informed Consent Statement: Informed consent was obtained from all subjects involved in the study.

Data Availability Statement: The data presented in this study are available on request from the corresponding author

Conflicts of Interest: The authors declare no conflicts of interest.

Acknowledgements: The authors are thankful to Dr. Yasuaki Koyama, MD, PhD for valuable suggestions.